# Deformation of an elastic body in low Reynolds number transport: Relevance to biofilm deformation and streamer formation

Nikhil Gupta[Y], Siddhartha Das[†], Sushanta K. Mitra[†,1], Aloke Kumar[†*]

[Y]Department of Mechanical Engineering, Indian Institute of Technology, Bombay, India

[†]Department of Mechanical Engineering, University of Alberta, Edmonton, AB Canada

*Corresponding author email: aloke.kumar@ualberta.ca

---

[1] Current address: Lassonde School of Engineering, York University, Toronto, ON, Canada

**Abstract**
In this paper, we obtain analytical results for shear stress distributions inside an elastic body placed in a low Reynolds number transport. The problem definition is inspired by a recent experimental study (Valiei et al., *Lab Chip*, 2012, **12**, 5133-5137) that reports the flow-triggered deformation of bacterial biofilms, formed on cylindrical rigid microposts, into long filamentous structures known as streamers. In our analysis, we consider an elastic body of finite thickness (forming a rim) placed over a rigid cylinder, i.e., we mimic the biofilm structure in the experiment. We consider Oseen flow solution to describe the low Reynolds transport past this cylindrical elastic structure. The stress and strain distributions inside the elastic structure are found to be functions of position, Poisson ratio, initial thickness of the elastic rim and the ratio of the flow-driven shear stress to the shear modulus of the elastic body. More importantly, these analyses, which can be deemed as one of the first formal analyses to understand the *fluid-structure-interaction* issues associated with the biofilm streamer formation, help us interpret several qualitative aspects associated with the streamer formation reported in different experiments.

# 1. Introduction

Understanding the flow-induced deformation of elastic or soft materials often becomes necessary to interpret different biophysical phenomena. These problems, often classified as fluid-structure interaction problems, can appear in both static [1-3] and flow conditions [4-6] spanning different decades of length and time scales. Some of the well-known examples in this regard are flagella driven motion of bacteria [7, 8], transport and deformation of mammalian cells [9, 10], cilia response to shear stresses [11, 12], etc. One such problem that is still very much in its infancy is the deformation behavior of bacterial biofilms in the presence of a background flow field [13]. Bacterial biofilms, which are the surface-hugging aggregates of microbes encased by extracellular polymeric substances (EPS) and often considered to be indispensable in the understanding of bacterial activities in natural habitats, have been widely studied in context of diverse applications ranging from human health to environment [14-19]. Cells encased in biofilms show markedly different behaviors as compared to the non-biofilm-encapsulated bacteria grown as planktonic cells in the laboratory. Ubiquity of biofilms and their broad relevance to human activities have prompted a major upsurge in developing lab scale experiments that can focus on the dynamics of biofilm formation – such formation occurs spontaneously in a bacterial solution with exposed solid-liquid interfaces – and understand the impact of external stimuli, such as an applied electric field [20], or a background flow field [21]. Biofilms in practical scenarios can often be subject to background flows, such as flows over heart stents and even flows over ship hulls [22, 23]. Presence of a flow field can elicit different responses from a biofilm ranging from simple deformation to fracture. But, as has been reported in several recent experiments, the more intriguing response of the biofilms to an applied flow field, with the corresponding *Reynolds number* associated with the flow field being much smaller than unity, is their 'extrusion' into long, slender filamentous structures known as streamers [24-27]. Streamer formation can have wide repercussions including rapid clogging of devices [23] as well as substantial fluid-structure interactions [28, 29]. One of the few analytical studies that focuses on fluid-structure interaction with respect to streamers is the study by Autrusson et al. [30], where the authors attempt to understand streamer shape by considering the dynamics of an elastic filament subject to flow induced shear. While this study is an important milestone in understanding the post-formation dynamics of streamers, the stress picture to which a pre-formed biofilm is subjected to prior to streamer formation remains mostly unaddressed. In another study, Rusconi et al. [31] computed the flow field in the geometry where streamers are witnessed, and correlated the location from where streamers evolve with the corresponding flow behavior at such locations. However, to the study did not correlate the flow-imposed stresses to the corresponding elastic deformations through actual elastic stress-strain calculations. In this light, our present calculation becomes significant – it provides for the first time flow-induced stress-strain response on an elastic mass (i.e., our analysis provides a coupled flow field and elastic deformation calculation), which can be assumed to replicate a pre-formed biofilm prior to its 'extrusion' into a streamer. Such stress-strain calculation is fundamental to decipher the formation event (and the relevant timescale $t_f$) of the biofilm streamers, given the inevitable dependence of the process on the stress-strain history of the intrinsically viscoelastic biofilm.

Stress-strain calculations for biofilms are strongly dictated by the specific environment in which biofilms exist. Streamer formation has been witnessed in biofilms formed in various kinds of geometries, such as on cylindrical microposts [26], in the corners of a serpentine channel [25], at a horse-shoe shaped obstruction placed in the path of a fluid flow [27], etc. For our calculations, we consider the case of Valiei et al. [26], where biofilms are formed on cylindrical microposts. The central hypothesis employed here is that the biofilms exist as preformed elastic mass, and we study the stress-strain behavior of this elastic mass in response to a flow induced shear. Such a problem statement inevitably ignores the dynamic nature of the biofilm growth process; in other words the biofilm thickness is assumed to be constant and the flow is considered only to impart a mechanical stress on the biofilm.

In this paper, we obtain detailed analytical solutions for the flow-induced stress-strain behavior of a cylindrical elastic rim (see Fig. 1), replicating an elastic biofilm of finite constant thickness formed on the cylindrical microposts described in Valiei et al. [26]. The calculations invoke the Oseen solution to describe the flow field, and the Airy stress function method to calculate the resulting stress-strain behavior. The central result of the paper is that there is a significant position dependent variation of the shear stresses and strains, as well as variation with Poisson ratio ($\nu$) and the original thickness of the elastic rim. In fact, depending on these parameters we may get a stress distribution inside the elastic mass that does not trivially follow from the flow imparted stress at the boundary of the elastic mass and the background fluid. More importantly, the information about the stress distribution inside the elastic rim (replicating the biofilm) as a function of different system parameters provides most useful insights about the streamer formation event. Given the scarcity of systematic fluid-structure interaction studies intrinsic to biofilm streamer formation, our work will serve as an important starting point to develop the necessary theoretical understanding needed to explain the streamer formation.

## 2 Theory

### 2.1 Stresses exerted on the cylinder surface by the fluid flow

We start our analysis by calculating the stresses exerted on an infinite cylinder of radius $ap$ [covered with an elastic rim, see Fig. 1(a)] by a fluid flowing past it [see Fig. 1(a)]. We consider the free stream velocity $u_\infty$ to be substantially small (atypical of most of the microfluidic systems), so that we can invoke the *Oseen solution* for expressing the flow past a cylinder in a *creeping flow* [or low Reynolds number ($Re$)] scenario. Oseen derived the governing equations for the flow field, whereas the actual analytical solution was obtained by Lamb [32]. Here we shall only state the key results governing flow field. Interested readers may kindly look into the available literature [32] for the method of derivation and solution of the equations. The two-dimensional velocity field (i.e., velocity field with components along $r$ and $\theta$ directions) can be expressed in dimensionless form as:

$$U_r = \cos\theta + \frac{\partial \varphi}{\partial R} + \Pi \cos\theta - \frac{1}{2q}\frac{\partial \Pi}{\partial R}, \tag{1}$$

$$U_\theta = -\sin\theta + \frac{1}{R}\frac{\partial \varphi}{\partial \theta} - \Pi \sin\theta - \frac{1}{2qR}\frac{\partial \Pi}{\partial \theta}. \tag{2}$$

In the above equations, $U_r = u_r/u_\infty$ (where $u_r$ is the velocity in $r$ direction) and $U_\theta = u_\theta/u_\infty$ (where $u_\theta$ is the velocity in $\theta$ direction) are the dimensionless velocities in $r$ and $\theta$ directions, respectively. Also $R=r/a$ [$a$ is the radius of the cylinder including the elastic rim, see Fig. 1(a)] is the dimensionless radial coordinates and $q = Re/4$ (where $Re=2a\rho u_\infty/\mu$ is the Reynolds number; $\rho$ and $\mu$ are the density and the dynamic viscosity of the liquid). Further the functions $\varphi$ and $\Pi$ can be expressed as:

$$\varphi = A \ln R + \frac{B}{R}\cos\theta, \tag{3}$$

$$\Pi = Ce^{qR\cos\theta} K_0(qR), \tag{4}$$

where $A$, $B$ and $C$ are the constants to be determined from the boundary conditions (on the flow field) and $K_0$ is the modified Bessel function of second kind of order zero. These boundary conditions are the no slip and the no penetration conditions at the elastic rim surface, i.e.,

$$(U_\theta)_{R=1} = 0 \text{ (No slip condition)}, \quad (U_R)_{R=1} = 0 \text{ (No penetration condition)}. \tag{5}$$

Using eqs.(3,4) in eqs.(1,2) and imposing the conditions of eq.(5) we get the constants [appearing in eqs.(3,4)] as:

$$A = -\frac{1}{\varepsilon}\frac{1}{\left[\alpha+\ln\left(\frac{q}{2}\right)\right]-\frac{1}{2}}, \quad B = \frac{\frac{1}{2}}{\left[\alpha+\ln\left(\frac{q}{2}\right)\right]-\frac{1}{2}}, \quad C = \frac{2}{\left[\alpha+\ln\left(\frac{q}{2}\right)\right]-\frac{1}{2}}. \tag{6}$$

To arrive at eq.(6), we additionally use the condition $qR \ll 1$, so that we can simplify eq.(4) by using expanded forms of the Bessel function [i.e., $K_0(qR) \approx -\{\alpha+\ln(qR/2)\}$, where $\alpha=0.57$ is the Euler's constant] and the exponential function (i.e., $e^{qR\cos\theta} \approx 1+qR\cos\theta$).

Once the velocity field has been obtained [see eqs.(1-4,6)], we can calculate the strain rates (we denote the corresponding dimensional form as $\varepsilon_{ij}$ and the dimensionless form as $\bar\varepsilon_{ij}$) in the flow field as:

$$\bar\varepsilon_{rr} = \frac{\varepsilon_{rr}}{(u_\infty/a)} = \frac{\partial U_r}{\partial R}, \quad \bar\varepsilon_{\theta\theta} = \frac{\varepsilon_{\theta\theta}}{(u_\infty/a)} = \frac{1}{R}\left(\frac{\partial U_\theta}{\partial \theta} - U_r\right), \quad \bar\varepsilon_{r\theta} = \frac{\varepsilon_{r\theta}}{(u_\infty/a)} = \frac{1}{R}\left(\frac{\partial U_r}{\partial \theta} - U_\theta\right) + \frac{\partial U_\theta}{\partial R}. \tag{7}$$

From the strain rates, we can get the stresses (we denote the corresponding dimensional form as $\sigma_{ij}$ and the dimensionless form as $\bar\sigma_{ij}$) inside the flow field as:

$$\bar\sigma_{rr} = \frac{\sigma_{rr}}{(\mu u_\infty/a)} = -\frac{P_0}{(\mu u_\infty/a)} + 2\bar\varepsilon_{rr}, \quad \bar\sigma_{\theta\theta} = \frac{\sigma_{\theta\theta}}{(\mu u_\infty/a)} = 2\bar\varepsilon_{\theta\theta}, \quad \bar\sigma_{r\theta} = \frac{\sigma_{r\theta}}{(\mu u_\infty/a)} = \bar\varepsilon_{r\theta}, \tag{8}$$

where $P_0$ is the pressure and $\mu$ is the dynamic viscosity.

Therefore, the stresses exerted at the elastic rim boundary [these stresses (with a negative sign) will be later used as the boundary condition for solving the stress distribution inside the elastic rim] are:

$$(\bar{\sigma}_{rr})_{R=1} = -(A_1 \cos 2\theta + B_1 \cos \theta + C_1), \quad (\bar{\sigma}_{\theta\theta})_{R=1} = -(A_2 \cos 2\theta + B_2 \cos \theta + C_2),$$

$$(\bar{\sigma}_{r\theta})_{R=1} = -(A_3 \sin 2\theta + B_3 \sin \theta), \tag{9}$$

where

$$A_1 = -C\alpha q - Cq - Cq \ln(q/2), \tag{10a}$$

$$B_1 = -C - 2Bq + 4B, \tag{10b}$$

$$C_1 = 2qA + \frac{A}{2q} + \frac{C}{q} - C\alpha q - Cq - Cq \ln(q/2), \tag{10c}$$

$$C_2 = Cq[\alpha + \ln(q/2)] - \frac{C}{q} - 2A, \tag{10d}$$

$$A_2 = 3Cq[\alpha + \ln(q/2)], \tag{10e}$$

$$B_2 = -4 - C + \frac{7C}{2}[\alpha + \ln(q/2)], \tag{10f}$$

$$A_3 = \frac{3Cq}{2} + Cq \ln(q/2), \tag{10g}$$

$$B_3 = 4B. \tag{10h}$$

## 2.2 Deformations of the elastic rim

We are considering an infinitely long cylindrical body, whose outer layer is an elastic rim (thickness of this elastic rim is $a-ap$), whereas the inner layer is a perfectly rigid cylinder [see Fig. 1(a)]. We assume a condition of plane strain, i.e., the strains are present only in the $r$-$\theta$ plane and there are no strains in $z$ direction. Such an assumption is routinely applied in calculation of elastic deformation of an infinite (in $z$) cylinder [33]. Under this situation, we can express the displacements (we denote the corresponding dimensional form as $x_i$ and the dimensionless form as $\bar{x}_i$) as [34]:

$$\bar{x}_r = \frac{x_r}{a} = \bar{x}_r(R,\theta), \quad \bar{x}_\theta = \frac{x_\theta}{a} = \bar{x}_\theta(R,\theta), \quad \bar{x}_z = \frac{x_z}{a} = 0, \tag{11}$$

where $x_r$, $x_\theta$ and $x_z$ are the displacements (dimensional) in $r$, $\theta$ and $z$ directions. Further we can relate the displacements to the corresponding strains $\bar{e}_{ij}$ (which are dimensionless) as:

$$\bar{e}_{rr} = \frac{\partial \bar{x}_r}{\partial R}, \quad \bar{e}_{\theta\theta} = \frac{1}{R}\left(\frac{\partial \bar{x}_\theta}{\partial \theta} + \bar{x}_r\right), \quad \bar{e}_{r\theta} = \frac{1}{2}\left(\frac{1}{R}\frac{\partial \bar{x}_r}{\partial \theta} + \frac{\partial \bar{x}_\theta}{\partial R} - \frac{\bar{x}_\theta}{R}\right). \tag{12}$$

We would now like to express the strains ($\bar{e}_{ij}$) in terms of the corresponding dimensionless stresses ($\bar{\sigma}_{ij}$) using Hooke's Law [34]:

$$\bar{e}_{zz} = \left(\frac{\mu u_\infty}{aE}\right)\left[\bar{\sigma}_{zz} - \nu(\bar{\sigma}_{rr} + \bar{\sigma}_{\theta\theta})\right] = 0 \text{ (Plane strain condition)}$$

$$\Rightarrow \bar{\sigma}_{zz} = \nu(\bar{\sigma}_{rr} + \bar{\sigma}_{\theta\theta}),$$

$$\bar{e}_{rr} = \left(\frac{\mu u_\infty}{aE}\right)\left[\bar{\sigma}_{rr} - \nu(\bar{\sigma}_{\theta\theta} + \bar{\sigma}_{zz})\right] \Rightarrow \bar{e}_{rr} = \left(\frac{\mu u_\infty}{aE}\right)(1+\nu)\left[(1-\nu)\bar{\sigma}_{rr} - \nu\bar{\sigma}_{\theta\theta}\right], \quad (13)$$

$$\bar{e}_{\theta\theta} = \left(\frac{\mu u_\infty}{aE}\right)\left[\bar{\sigma}_{\theta\theta} - \nu(\bar{\sigma}_{rr} + \bar{\sigma}_{zz})\right] \Rightarrow \bar{e}_{\theta\theta} = \left(\frac{\mu u_\infty}{aE}\right)(1+\nu)\left[(1-\nu)\bar{\sigma}_{\theta\theta} - \nu\bar{\sigma}_{rr}\right],$$

$$\bar{e}_{r\theta} = \left(\frac{\mu u_\infty}{aG}\right)(1+\nu)\bar{\sigma}_{r\theta},$$

where $E$ and $G$ are the elastic and shear moduli of the elastic rim present around the infinitely long rigid cylinder.

On the other hand, the force balance (or the equilibrium conditions) in dimensionless forms in $r$ and $\theta$ directions will yield [34]:

$$\frac{\partial \bar{\sigma}_{rr}}{\partial R} + \frac{1}{R}\frac{\partial \bar{\sigma}_{r\theta}}{\partial \theta} + \frac{(\bar{\sigma}_{rr} - \bar{\sigma}_{\theta\theta})}{R} + \bar{F}_r = 0, \quad (14)$$

$$\frac{\partial \bar{\sigma}_{r\theta}}{\partial R} + \frac{1}{R}\frac{\partial \bar{\sigma}_{\theta\theta}}{\partial \theta} + \frac{2}{R}\bar{\sigma}_{r\theta} + \bar{F}_\theta = 0, \quad (15)$$

where $\bar{F}_r = F_r/(\mu u_\infty/a^2)$ and $\bar{F}_\theta = F_\theta/(\mu u_\infty/a^2)$ are the dimensionless forms of the applied body forces (per unit volume) in $r$ and $\theta$ directions, respectively. For the present case, we have $\bar{F}_r = \bar{F}_\theta = 0$. Also, we can express the stresses in terms of the *Airy Stress Function* $\phi$ (since we are considering a plane strain problem [34]), such that (in absence of any external force):

$$\bar{\sigma}_{rr} = \frac{1}{R}\frac{\partial \phi}{\partial R} + \frac{1}{R^2}\frac{\partial^2 \phi}{\partial \theta^2}, \quad \bar{\sigma}_{\theta\theta} = \frac{\partial^2 \phi}{\partial R^2}, \quad \bar{\sigma}_{r\theta} = -\frac{\partial}{\partial R}\left(\frac{1}{R}\frac{\partial \phi}{\partial \theta}\right). \quad (16)$$

Such representation of the stresses in terms of $\phi$ identically satisfies the equilibrium conditions [eqs. (14,15)], and at the same time ensures that the problem can be described in terms of a single governing equation, involving a single variable (i.e., $\phi$). To obtain this key equation, we first express the strains through the compatibility relation:

$$\frac{1}{R^2}\frac{\partial^2 \bar{e}_{rr}}{\partial \theta^2} + \frac{\partial^2 \bar{e}_{\theta\theta}}{\partial R^2} - \frac{2}{R}\frac{\partial^2 \bar{e}_{r\theta}}{\partial R \partial \theta} - \frac{1}{R}\frac{\partial \bar{e}_{rr}}{\partial R} + \frac{2}{R}\frac{\partial \bar{e}_{\theta\theta}}{\partial R} - \frac{2}{R^2}\frac{\partial \bar{e}_{r\theta}}{\partial \theta} = 0. \quad (17)$$

We use eq.(16), in eq.(13) to express the strain components $\bar{e}_{ij}$ in terms of $\phi$, and then substitute these strains in eq.(17) to obtain this single equation governing the problem as:

$$\left(\frac{\partial^2}{\partial R^2} + \frac{1}{R}\frac{\partial}{\partial R} + \frac{1}{R^2}\frac{\partial^2}{\partial \theta^2}\right)^2 \phi = \nabla^4 \phi == 0. \qquad (18)$$

Therefore the Airy Stress Function $\phi$ is governed by a biharmonic equation. Following [34], we can obtain a general solution for $\phi$ as:

$$\begin{aligned}\phi &= a_0 + a_1 \log R + a_2 R^2 + a_3 R^2 \log R + \left[a_4 + a_5 \log R + a_6 R^2 + a_7 R^2 \log R\right]\theta \\ &+ \left(a_{11}R + a_{12}R\log R + \frac{a_{13}}{R} + a_{14}R^3 + a_{15}R\theta + a_{16}R\log R\theta\right)\cos\theta \\ &+ \left(b_{11}R + b_{12}R\log R + \frac{b_{13}}{R} + b_{14}R^3 + b_{15}R\theta + b_{16}R\log R\theta\right)\sin\theta \\ &+ \sum_{n=2}^{\infty}\left(a_{n1}r^n + a_{n2}r^{2+n} + a_{n3}r^{-n} + a_{n4}r^{2-n}\right)\cos n\theta \\ &+ \sum_{n=2}^{\infty}\left(b_{n1}r^n + b_{n2}r^{2+n} + b_{n3}r^{-n} + b_{n4}r^{2-n}\right)\sin n\theta,\end{aligned} \qquad (19)$$

where $a_i$ ($i=0,1...,7$), $a_{1j}$ ($j=1,2....,6$), $b_{1j}$, $a_{nk}$ ($k=1,2,3,4$) and $b_{n,k}$ are all constants. These constants need to be evaluated from the boundary conditions. We have two sets of boundary conditions – the first set of boundary conditions is at $R=1$ and expresses the stresses imparted by the creeping flow, whereas the second set of boundary condition expresses a physical constraint imposed at the inner boundary (i.e., $R=p$). These boundary conditions are [see eq.(9)]:

$$\begin{aligned}(\bar{\sigma}_{rr})_{R=1} &= A_1 \cos 2\theta + B_1 \cos\theta + C_1 \\ (\bar{\sigma}_{\theta\theta})_{R=1} &= A_2 \cos 2\theta + B_2 \cos\theta + C_2 \\ (\bar{\sigma}_{r\theta})_{R=1} &= A_3 \sin 2\theta + B_3 \sin\theta,\end{aligned} \qquad (20)$$

$$(\bar{x}_r)_{R=p} = 0 \Rightarrow \left(\int \bar{\varepsilon}_{rr} dR\right)_{R=p} = 0. \qquad (21)$$

Eq.(21) represents the physical constraint of *no radial displacement* at the inner boundary (i.e., the interface between the elastic and the rigid body), and the corresponding strain condition is expressed under the assumption that there is no rigid-body translation and rotation (i.e., no displacement in absence of an imposed strain). Also the nature of the transcendental boundary conditions expressed in eq. (20) allows us to simplify the generalized solution for $\phi$ [see eq.(19)] as:

$$\phi = a_0 + a_1 \log R + a_2 R^2 + a_3 R^2 \log R + \left[ a_4 + a_5 \log R + a_6 R^2 + a_7 R^2 \log R \right] \theta$$

$$+ \left( a_{11} R + a_{12} R \log R + \frac{a_{13}}{R} + a_{14} R^3 + a_{15} R\theta + a_{16} R \log R\theta \right) \cos \theta$$

$$+ \left( b_{11} R + b_{12} R \log R + \frac{b_{13}}{R} + b_{14} R^3 + b_{15} R\theta + b_{16} R \log R\theta \right) \sin \theta \tag{22}$$

$$+ \left( a_{21} R^2 + a_{22} R^4 + \frac{a_{23}}{R^2} + a_{24} \right) \cos 2\theta + \left( b_{21} R^2 + b_{22} R^4 + \frac{b_{23}}{R^2} + b_{24} \right) \sin 2\theta.$$

Solving eq. (22) in presence of eqs. (20,21) [we invoke eqs. (13,16) to express the boundary conditions in terms of $\phi$], we finally get the different constants as follow:

$$a_1 = C_1 \left( \frac{p^2 \log p + 3 p^2 \nu - 2 p^2 - 2 \nu p^2 \log p}{3 p^2 - 4 p^2 \nu - 2 p^2 \log p + 4 \nu p^2 \log p + 1} \right)$$

$$- C_2 \left( \frac{p^2 \log p + p^2 \nu - p^2 - 2 p^2 \nu \log p}{3 p^2 - 4 p^2 \nu - 2 p^2 \log p + 4 \nu p^2 \log p + 1} \right), \tag{23a}$$

$$a_2 = -\frac{C_1}{4} \left( \frac{p^2 - 2 p^2 \log p + 4 p^2 \nu \log p + 3}{3 p^2 - 4 p^2 \nu - 2 p^2 \log p + 4 \nu p^2 \log p + 1} \right)$$

$$+ \frac{C_2}{4} \left( \frac{2 p^2 \log p - p^2 - 4 \nu p^2 \log p + 1}{3 p^2 - 4 p^2 \nu - 2 p^2 \log p + 4 \nu p^2 \log p + 1} \right), \tag{23b}$$

$$a_3 = \frac{C_1}{2} \left( \frac{2 p^2 \nu - p^2 + 1}{3 p^2 - 4 p^2 \nu - 2 p^2 \log p + 4 \nu p^2 \log p + 1} \right)$$

$$- \frac{C_2}{2} \left( \frac{p^2 - 2 p^2 \nu + 1}{3 p^2 - 4 p^2 \nu - 2 p^2 \log p + 4 \nu p^2 \log p + 1} \right), \tag{23c}$$

$$a_4 = a_5 = a_6 = a_7 = a_{11} = 0, \tag{23d}$$

$$a_{12} = 4 B_1 \left( \frac{p^2 \log p - p^2 \nu \log p}{4 p^2 \log p + 4 p^4 \nu - p^4 - 8 p^2 \nu \log p + 1} \right) +$$

$$\frac{B_2}{2} \left( \frac{p^4 - 4 p^4 \nu + 1}{4 p^2 \log p + 4 p^4 \nu - p^4 - 8 p^2 \nu \log p + 1} \right) - \frac{B_3}{2} \left( \frac{8 p^2 \log p + 4 p^4 \nu - p^4 - 8 p^2 \nu \log p + 3}{4 p^2 \log p + 4 p^4 \nu - p^4 - 8 p^2 \nu \log p + 1} \right), \tag{23e}$$

$$a_{13} = B_1 \left( \frac{p^2 \log p - p^2 \nu \log p}{4 p^2 \log p + 4 p^4 \nu - p^4 - 8 p^2 \nu \log p + 1} \right) -$$

$$\frac{B_2}{4} \left( \frac{2 p^2 \log p + 4 p^2 \nu - p^4 - 4 p^2 \nu \log p}{4 p^2 \log p + 4 p^4 \nu - p^4 - 8 p^2 \nu \log p + 1} \right) + \frac{B_3}{4} \left( \frac{2 p^2 \log p + 4 p^4 \nu - p^4 - 8 p^2 \nu \log p}{4 p^2 \log p + 4 p^4 \nu - p^4 - 8 p^2 \nu \log p + 1} \right), \tag{23f}$$

$$a_{14} = -B_1\left(\frac{p^2\log p - p^2\nu\log p}{4p^2\log p + 4p^4\nu - p^4 - 8p^2\nu\log p + 1}\right) - $$
$$\frac{B_2}{4}\left(\frac{2p^2\log p - 4p^2\nu\log p + 1}{4p^2\log p + 4p^4\nu - p^4 - 8p^2\nu\log p + 1}\right) + \frac{B_3}{4}\left(\frac{2p^2\log p + 1}{4p^2\log p + 4p^4\nu - p^4 - 8p^2\nu\log p + 1}\right),\quad(23g)$$

$$a_{15} = a_{16} = b_{11} = b_{12} = b_{13} = b_{14} = b_{16} = 0, \quad(23h)$$

$$b_{15} = \frac{B_3 - B_1}{2}, \quad(23i)$$

$$a_{21} = \frac{A_1}{4}\left(\frac{9p^2\nu + p^6\nu - 9p^2 + 1}{6p^2\nu + 2p^6\nu - 6p^2 - 3p^4 + 1}\right) + $$
$$\frac{A_2}{4}\left(\frac{3p^2\nu - p^6\nu - 3p^2 + 1}{6p^2\nu + 2p^6\nu - 6p^2 - 3p^4 + 1}\right) - \frac{A_3}{2}\left(\frac{3p^2\nu + p^6\nu - 3p^2 + 1}{6p^2\nu + 2p^6\nu - 6p^2 - 3p^4 + 1}\right), \quad(23j)$$

$$a_{22} = -\frac{A_1}{24}\left(\frac{12p^2\nu - 12p^2 + 3p^4 + 1}{6p^2\nu + 2p^6\nu - 6p^2 - 3p^4 + 1}\right) - \frac{A_2}{8}\left(\frac{4p^2\nu - 4p^2 - p^4 + 1}{6p^2\nu + 2p^6\nu - 6p^2 - 3p^4 + 1}\right)$$
$$+ \frac{A_3}{12}\left(\frac{3p^4 + 1}{6p^2\nu + 2p^6\nu - 6p^2 - 3p^4 + 1}\right), \quad(23k)$$

$$a_{23} = \frac{A_1}{12}\left(\frac{3p^2\nu - p^6\nu - 3p^2 + 3p^4}{6p^2\nu + 2p^6\nu - 6p^2 - 3p^4 + 1}\right) - \frac{A_2}{4}\left(\frac{p^2\nu + p^6\nu - p^2 - p^4}{6p^2\nu + 2p^6\nu - 6p^2 - 3p^4 + 1}\right)$$
$$+ \frac{A_3}{6}\left(\frac{3p^2\nu + p^6\nu - 3p^4}{6p^2\nu + 2p^6\nu - 6p^2 - 3p^4 + 1}\right), \quad(23l)$$

$$a_{24} = \frac{A_1}{8}\left(\frac{4p^6\nu - 9p^4 + 1}{6p^2\nu + 2p^6\nu - 6p^2 - 3p^4 + 1}\right) - \frac{A_2}{8}\left(\frac{-4p^6\nu + 3p^4 + 1}{6p^2\nu + 2p^6\nu - 6p^2 - 3p^4 + 1}\right)$$
$$+ \frac{A_3}{4}\left(\frac{3p^4 + 1}{6p^2\nu + 2p^6\nu - 6p^2 - 3p^4 + 1}\right), \quad(23m)$$

$$b_{21} = b_{22} = b_{23} = b_{24} = 0. \quad(23n)$$

Once eq.(22) has been solved, we can use eq.(16) to obtain the stresses $\bar{\sigma}_{rr}$, $\bar{\sigma}_{\theta\theta}$ and $\bar{\sigma}_{r\theta}$.

## 3 Results

### 3.1 Variation of the flow field

Fig. 1(b) shows the *top-view* of the velocity vectors around the vertical cylinder. The no-slip and the no-penetration boundary conditions at the elastic rim surface (i.e., outer surface of the elastic rim) are clearly visible the notable influence of the cylinder in altering the flow field in its vicinity. This result, as well as those on elastic calculations that will follow, are provided for a substantially small *Reynolds number* (Re) of $10^{-3}$, which allows us to invoke the Oseen solution [31] to obtain the flow field. The most important issue here is to understand the variation of the velocity vectors in the vicinity of the elastic rim surface ($R \to 1^+$), since that dictates the stresses imparted on the elastic mass. First, we find that for $\theta=0, \pi, 2\pi$, in the region $R \to 1^+$, the tangential velocities ($U_\theta$) are non-existent, whereas the radial velocities ($U_r$) are very small and show negligible variation with $\theta$. This would imply that the shear strain rates in the liquid at these locations vanish, i.e., $\left[ \bar{\varepsilon}_{r\theta} = \frac{1}{R}\left( \frac{\partial U_r}{\partial \theta} - U_\theta \right) + \frac{\partial U_\theta}{\partial R} \right]_{R \to 1^+; \theta=0, \pi, 2\pi} \to 0$, ensuring that the corresponding stresses exerted by the fluid on the elastic solid at these locations also vanish (see later). In fact for any value of $\theta$, at $R \to 1^+$, we always have $U_\theta = 0$ (due to no slip condition) and $\frac{\partial U_r}{\partial \theta} \to 0$ (due to no penetration condition for all $\theta$), which will imply $\left( \bar{\varepsilon}_{r\theta} \right)_{R \to 1^+; \forall \theta} \approx \frac{\partial U_\theta}{\partial R}$. Since we are using Oseen solution to describe the flow field, we recover the free stream condition far away from the elastic rim surface, and this ensures that $U_\theta$ increases from very small value (at $R \to 1^+$) rapidly away from elastic rim surface to attain the free stream value. Please note that $\left( \frac{\partial U_\theta}{\partial R} \right)_{R \to 1^+}$ and hence the flow-imparted stress will be largest corresponding to those $\theta$ values for which $U_\theta$ is largest in the far stream. Such locations are $\theta \to \frac{\pi^\pm}{2}, \frac{3\pi^\pm}{2}$ (see later). Accordingly, at these $\theta$ values on the elastic rim surface we get the highest magnitude of the flow-imparted stress. This analysis signifies the importance of using Oseen solution in describing the flow field – this approach allows to recover the free stream condition far away from the elastic rim surface (this is not possible with Stokes solution), and therefore leads to an accurate prediction of $\frac{\partial U_\theta}{\partial R}$ and the flow-imparted shear stress at the surface of the elastic rim.

### 3.2 Stress Distribution within the elastic cylinder

Fig. 2 shows the variation of the shear stress ($\bar{\sigma}_{r\theta}$) inside the elastic rim as a function of $\theta$ and $R$ for different values of thickness of the elastic rim (characterized by $p$; larger $p$ implies smaller thickness of the rim) and the Poisson ratio ($\nu$). Irrespective of $p$, $\nu$ and $R$ (for a given $p$) values, we always find zero shear stress at $\theta=0, \pi, 2\pi$, owing to zero imposed shear stress by the fluid flow for such $\theta$ values (see above). Also for all the $p$ values in the incompressible regime ($\nu=0.49$), as well as large $p$ values (e.g., $p=0.8$) for smaller $\nu$ ($=0.1$), we find that for a given radial location inside the elastic rim, the shear stresses show maximum in the $\theta \to \frac{\pi^{\pm}}{2}, \frac{3\pi^{\pm}}{2}$ range, being commensurate with the nature of the flow-induced stresses at the elastic rim surface (see above). Even for small $p$ (e.g. $p=0.2$) and small $\nu$ (=0.1) values, for most of radial locations, except for very small radial locations (i.e., locations in the vicinity of the interface between the elastic and the non-deformable solid), we find this behavior. Therefore, for all these radial locations (irrespective of the $p$ and $\nu$ values), we get a periodic behavior (with $\theta$) of the shear stress, where the shear starts from 0 at $\theta=0$, increases to a maximum for $\theta \to \frac{\pi^{\pm}}{2}$, then decreases again to 0 at $\theta=\pi$. This same behavior is repeated (with a negative sign) for the range $\theta=\pi$ to $\theta=2\pi$. For the cases where this standard periodic behavior is encountered, we find that for a given $p$ and $\theta$, as $R$ decreases (i.e., we come closer to the interface between the elastic and the non-deformable solid), shear stress increases. This is caused by the combination of the conditions at the boundaries ($R=1$, $R=p$). At the elastic rim surface ($R=1$), we have the condition of flow-imposed shear stress, whereas at the interface between the elastic and non-deformable solid ($R=p$), we have no radial displacement condition ($\bar{x}_r = 0$). Combination of these two factors lead to a stress build up inside the elastic rim, with the build up being most prominent at $R=p$, manifested by the highest magnitude of shear stress (for a given $p$, $\nu$ and $\theta$) at $R=p$. Since the shear stress at the external boundary ($R=1$) is specified, decrease in $p$ (or equivalently increase in thickness of the elastic rim) would mean that at $R=p$ (for smaller $p$) stresses are even more magnified. Magnification of the stresses at $R=p$ is even more substantial for the incompressible limit ($\nu=0.49$). Behavior at this limit can be understood by noting the variation of the corresponding strain $\bar{e}_{r\theta} = \frac{1}{2}\left(\frac{1}{R}\frac{\partial \bar{x}_r}{\partial \theta} + \frac{\partial \bar{x}_\theta}{\partial R} - \frac{\bar{x}_\theta}{R}\right)$. At $R=p$, $\bar{x}_r \equiv 0$, so that $\frac{\partial \bar{x}_r}{\partial \theta} = 0$, and hence (by incompressibility requirement) $\frac{\partial \bar{x}_\theta}{\partial R} = 0$, implying $\left(\bar{e}_{r\theta}\right)_{R=p} = -\frac{1}{2}\frac{\bar{x}_\theta}{R}$. The fact that we simultaneously have $\bar{x}_r \equiv 0$ as well as the incompressibility condition, there will be a catastrophic increase in $\bar{x}_\theta$, leading to a large increase in $\left(\bar{e}_{r\theta}\right)_{R=p}$.

The above analysis does not hold for smaller radial locations at small $p$ and $\nu$ values (e.g., see the case for $R=0.2$, $p=0.2$ and $\nu=0.1$). For such cases at $R=p$, we still have $\bar{x}_r \equiv 0$ and hence $\frac{\partial \bar{x}_r}{\partial \theta} = 0$; but now $\frac{\partial \bar{x}_\theta}{\partial R} \neq 0$ (since there is no incompressibility condition). Hence, $\left(\bar{e}_{r\theta}\right)_{R=p} = \frac{1}{2}\left(\frac{\partial \bar{x}_\theta}{\partial R} - \frac{\bar{x}_\theta}{R}\right)$ is now dictated by the interplay of $\bar{x}_\theta$ as well as its radial variation. This interplay effectively introduces an additional periodic behavior in the range $\theta=0$ to $\pi$ as well as $\theta=\pi$ to $2\pi$, so that the resultant shear stress distribution shows substantial deviation from the periodic behavior witnessed for the cases described previously.

## 4. Discussions

### 4.1 Significance of the dimensionless parameters in context of streamer formation

From eq. (13), as well as Fig. 2, we can identify three dimensionless parameters that dictate the problem, namely $e_0 = \frac{\mu u_\infty}{aG}$, $p$ and $\nu$. While $e_0$ directly governs the strength of the strains that quantifies the deformations, parameters $p$ and $\nu$ primarily dictate the qualitative response, although in certain cases may subtly influence the quantification as well (e.g., see in Fig. 2 the large magnitude of the dimensionless stress values for $\nu=0.49$ and $p=0.2$). In context of streamer formation, we can get a fair idea about the significance of the flow-induced shear and its influence in deforming the biofilms as streamers by obtaining an estimate $e_0$ for different experiments reporting streamer formation (off course the relevant length scale $a$ changes with the change in the flow configuration). This estimate is provided in Table I

**Table I:** *Summary of the flow-driven strain $e_0$. We obtain $u_\infty$ from Q using $u_\infty=Q/A$ (where A is the area of the flow passage). Also we always consider the solution viscosity $\mu=10^{-3}$ Pa-s.*

| Experiment | Bacteria forming biofilm | Q (μL/hr) | $u_\infty$ (mm/s) | $a$ (μm) | $\mu u_\infty/a$ (Pa) | $G$ (Pa) | $e_0 = \frac{\mu u_\infty}{aG}$ |
|---|---|---|---|---|---|---|---|
| Vallei et al. [26] | *Pseudomonas fluorescens* | ~ 10 | ~ 0.1 | Radius of micropost (~100 μm) | $10^{-3}$ | $10^{3}$ [35] | $10^{-4}$ |
| Rusconi et al. [25], Drescher et al. [23] | *Pseudomonas aeruginosa* | ~ 50 | ~ 1 | Microchannel dimension (~ 100 μm) | $10^{-2}$ | $10^{3}$ [35] | $10^{-3}$ |
| Weaver et al. [36] | *Staphylococcus Epidermis* | | | | 0.1–1 | $10^{3}$ [37] | $10^{-4} - 10^{-3}$ |

The strain values are low indicating a relatively small deformation of the biofilms, which ensures that the present linear model is an effective approximation. In fact, these strain values are important since they serve as the starting point to obtain the strain history (for the given shear stress from the imposed flow field) of the intrinsically viscoelastic biofilms, which would ultimately dictate their degeneration as streamers.

Poisson ratio $\nu$ dictates the role of the incompressibility effects in augmenting the stresses. Biofilms being mostly composed of EPS matrix that consists of polymers and proteins, we can consider a value of $\nu=0.45$ for the biofilms, as has been done elsewhere [38]. Therefore, biofilms can be approximated to be close to incompressible material, allowing us to paint the relevant stress picture from that depicted in Fig. 2 for the case of ν=0.49.

From the stress picture corresponding to the incompressible case, it is extremely important to quantify the *p* (or the biofilm thickness) value, since variation in *p* (for ν=0.49) can lead to as high as one order of magnitude increase in stress. For the streamer formation experiment of Valiei et al. [26], we find that *p* is mostly such that the biofilm thickness (when it starts to deform into streamers) is much smaller than the micropost radius, and therefore we can safely conclude that the strains mostly scale as $e_0$.

### 4.2 Relevance of stress distribution in context of streamer formation

In this section, we shall attempt to relate the stress picture in Fig. 2 to the different scenarios encountered in biofilm and streamer dynamics in presence of the applied flow field. For example, the radial dependence of the shear stress (for a given θ) for an incompressible solid, where the shear stress increases from the fluid-elastic-body interface to the elastic-body-rigid-solid interface has been reported elsewhere in context of effect of flow-induced shear on the deformation of biofilms [39]. We can also relate the θ-dependent variation of the shear stress (see Fig. 2) with the position (on the microposts) where the streamers first appear in the experiment of Valiei et al. [26]. As illustrated in Fig. 3(a), for the case corresponding to flow rate of 8 μL/hr, the streamers first appear at $\theta \to \frac{\pi^{\pm}}{2}$, and this nicely matches the location of the maximum shear stress for any *R* and *p* for the incompressible case [see Fig 3 (b)]. Please note that although our analysis (Fig. 2) is valid for elastic solid, and the deformation of the biofilms into streamers may be a more complicated viscoelastic effect, the above connection between the location of maximum shear [Fig. 3(b)] and the location from where the streamers start to appear [Fig. 3(a)] is still relevant. This is due to the fact that viscoelastic consideration introduces only time dependence in the stress-strain behavior, without changing the location of the maximum shear. At larger times, we do find streamers appearing from other angular (θ) locations (i.e., $\theta \neq \frac{\pi^{\pm}}{2}$) or the locations where shear stresses are not maximum. Also for larger flow rates, this time gap needed for streamers to appear from a θ location, such that $\theta \neq \frac{\pi^{\pm}}{2}$, is substantially reduced, ensuring a proliferation of the streamer density.

The results on the stress variation within the elastic rim (replicating the biofilm) as a function of *p* and ν (Fig. 2) can be used for more general interpretation of the streamer formation events, not necessarily captured by the experiment of Valiei et al. [26]. Certain details, e.g., values of θ where stresses become maximum pertain strictly to the geometry of the set up of Valiei et al. [26]. However, there are results that are independent of the experimental details of Valiei et al. [26]. For example, the result that for incompressible ν stresses (for a given θ) are maximum at *R=p* and this maximum increases with a decrease in *p* (i.e., increase in the biofilm thickness) is independent of the specifics of the set up of Valiei et al. [26], as has been validated by a numerical study in a completely different set up [39]. This finding can be extremely important for a particular streamer formation scenario that has not been discussed here. For example, usually the presence of a background flow not only exerts a mechanical shear to a preformed biofilm, but also at the same time adds mass to the biofilm, thereby increasing its thickness [26]. Streamer formation occurs in presence of the interplay of these two effects. Invoking our theoretical result that in the incompressible limit, thicker biofilm experiences almost an order-high magnitude of shear stress, such flow-induced mass addition effect must encounter a self-restraining effect, namely a catastrophic increase in the shear stress. This interpretation provided by our study points to a very important scenario where the streamer formation occurs by the mutually self-limiting influences of the fluid flow.

## 5 Conclusions

Through closed form analytical solution, we predict a *low Reynolds number* flow-imposed shear stress on circular elastic rim resting on a rigid cylinder. The geometry resembles the one employed by Valiei et al. [26] to study the flow-driven disintegration of biofilms into streamers. Therefore, our shear stress prediction gives important clues to issues such as the location from where streamers first start to develop, relative potential of different portions of the biofilm on a given structure to disintegrate into streamers etc. In addition, our theoretical model provides insights to streamer formation dynamics that are not even considered in the experiment of Valiei et al. [26]. Given the substantial rise in experiments that report and explain the significance of formation of streamers resulting from the response of a biofilm to a flow triggered shear, and a lack of analysis of *fluid-structure-interaction* effects responsible for streamer formation, our study will form an important starting point in making headway for a more quantitative analysis in this very challenging problem of biofilm streamer formation.

# Results

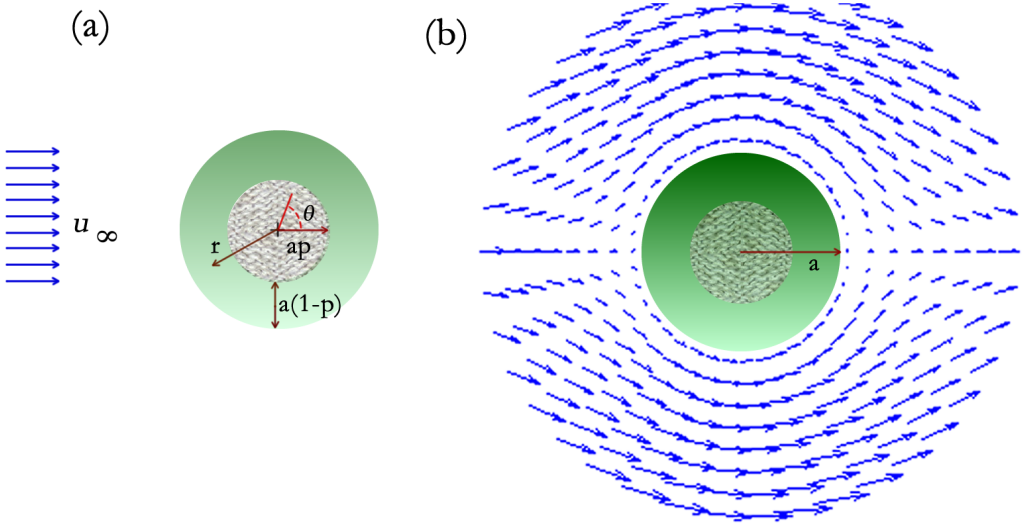

**Figure 1:** (a) Schematic of the problem (the elastic rim is shown in green and the rigid non-deformable cylinder is shown in gray). The lengths and the velocity are shown in (b) Velocity vector field around the elastic body.

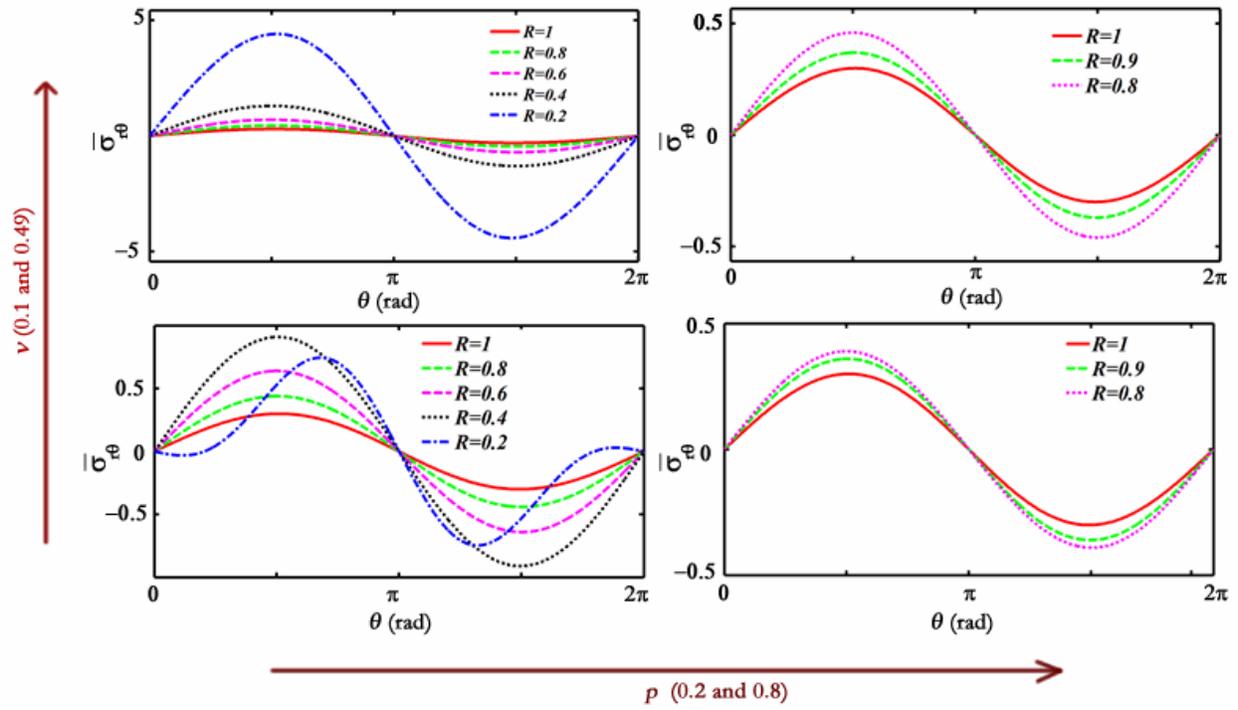

**Figure 2:** Variation of the shear stress inside the elastic body for different $\theta$ and radial ($R$) locations for different values of the Poisson ratio ($\nu$) and the initial rim thickness ($1 - p$).

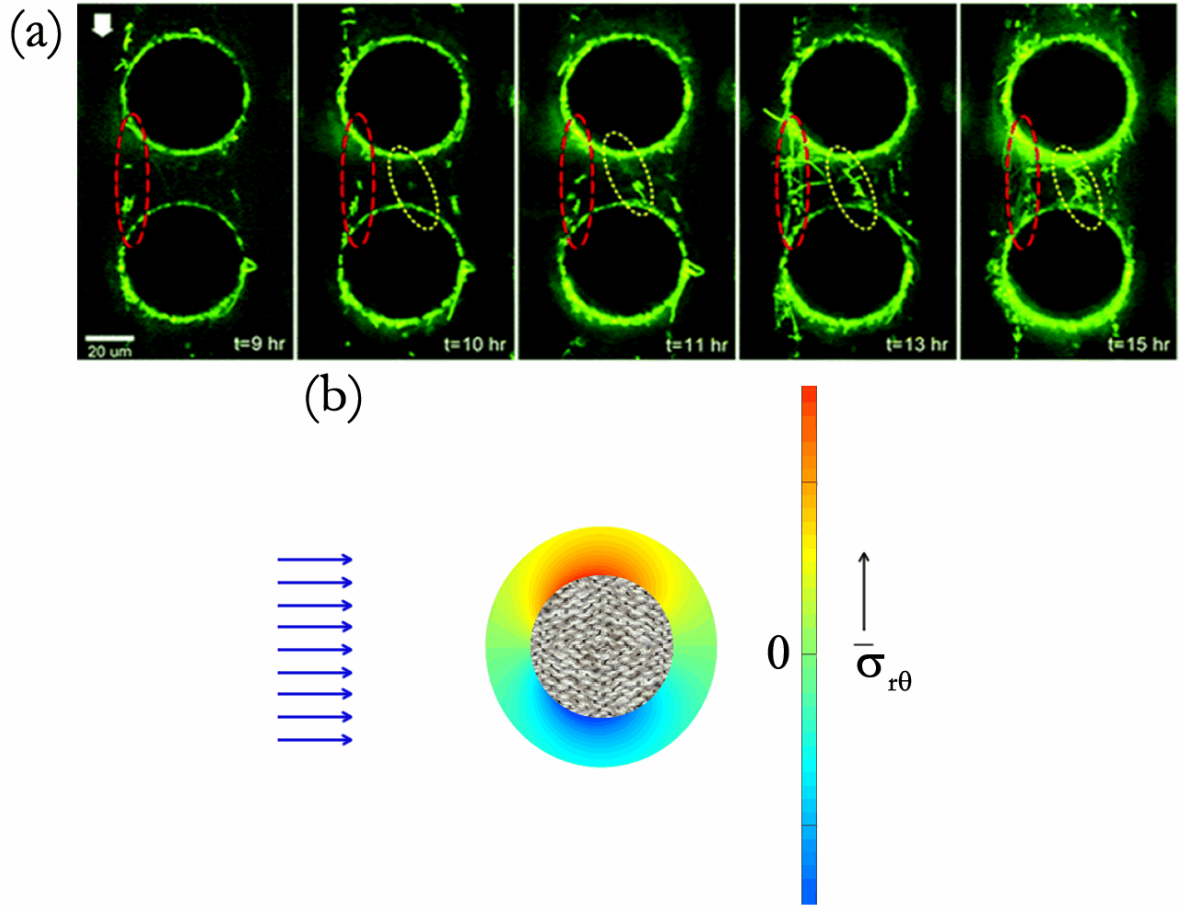

**Figure 3:** (a) Time evolution of streamer formation between two micro-pillars. Flow is from top to bottom. The green fluorescent protein (GFP) expressing bacteria are easily visualized against a dark background. Dashed ellipses show the position of streamers. The red ellipse shows a streamer forming near $\theta \rightarrow \frac{\pi^{\pm}}{2}$. Yellow ellipse shows a streamer at $\theta \neq \frac{\pi^{\pm}}{2}$. Valiei et al. [26] Reproduced by permission of The Royal Society of Chemistry. (b) Contour plot showing qualitative distribution of shear stress variation within the elastic rim.